\documentclass[ASTRA]{copernicus3}

\begin{document}

\title{Recent results from VHE gamma astrophysics related to fundamental physics and cosmic rays}

\author[]{Alessandro De Angelis}

\affil[]{Max-Planck-Institut f\"ur Physik
(Werner-Heisenberg-Institut),
F\"ohringer Ring 6,
M\"unchen, Germany\thanks{On leave of absence from Universit\`a di Udine, Via delle Scienze 208, Udine, Italy.};\\INFN and INAF Trieste, Italy; LIP/IST, Lisboa, Portugal
}


\runningtitle{VHE gamma astrophysics}

\runningauthor{A. De Angelis}

\correspondence{Alessandro De Angelis\\(alessandro.de.angelis@cern.ch)}

\received{}
\pubdiscuss{} 
\revised{}
\accepted{}
\published{}


\firstpage{1}

\abstract{High-energy photons are a powerful probe for astrophysics and for fundamental physics
under extreme conditions. During the recent years, our knowledge of the most violent
phenomena in the universe has impressively progressed thanks to the advent of new detectors
for very-high-energy (VHE) gamma rays (above 100 GeV). Ground-based detectors
like the Cherenkov telescopes (H.E.S.S., MAGIC and VERITAS) recently discovered
more than 80 new sources. This talk reviews the present status of VHE gamma astrophysics, with emphasis on the recent results and on the experimental
developments, keeping an eye on the future. The impact on fundamental physics
and on cosmic-ray physics is emphasized.}

\maketitle

\abstract{
 TEXT
 }
 
 \introduction

Despite the enormous success of Fermi, the fact that gamma-ray fluxes from astrophysical sources above the GeV have a typical energy dependence 
$E^{-\Gamma}$,
with the spectral index $\Gamma$ between 2 and 3, requires huge effective areas (of the order of $10^4-10^5$ m$^2$) to detect VHE (very-high energy, typically above 100 GeV) photons. Ground-based VHE gamma telescopes (such as MILAGRO, ARGO, H.E.S.S., MAGIC and VERITAS)
detect the secondary particles of the atmospheric showers produced by primary photons and 
cosmic rays (CR) of energy higher than the primaries observed by satellites.
Such ground-based detectors suffer a large amount of background from charged cosmic rays. 

\begin{figure}[htbp]
\includegraphics[width=\columnwidth]{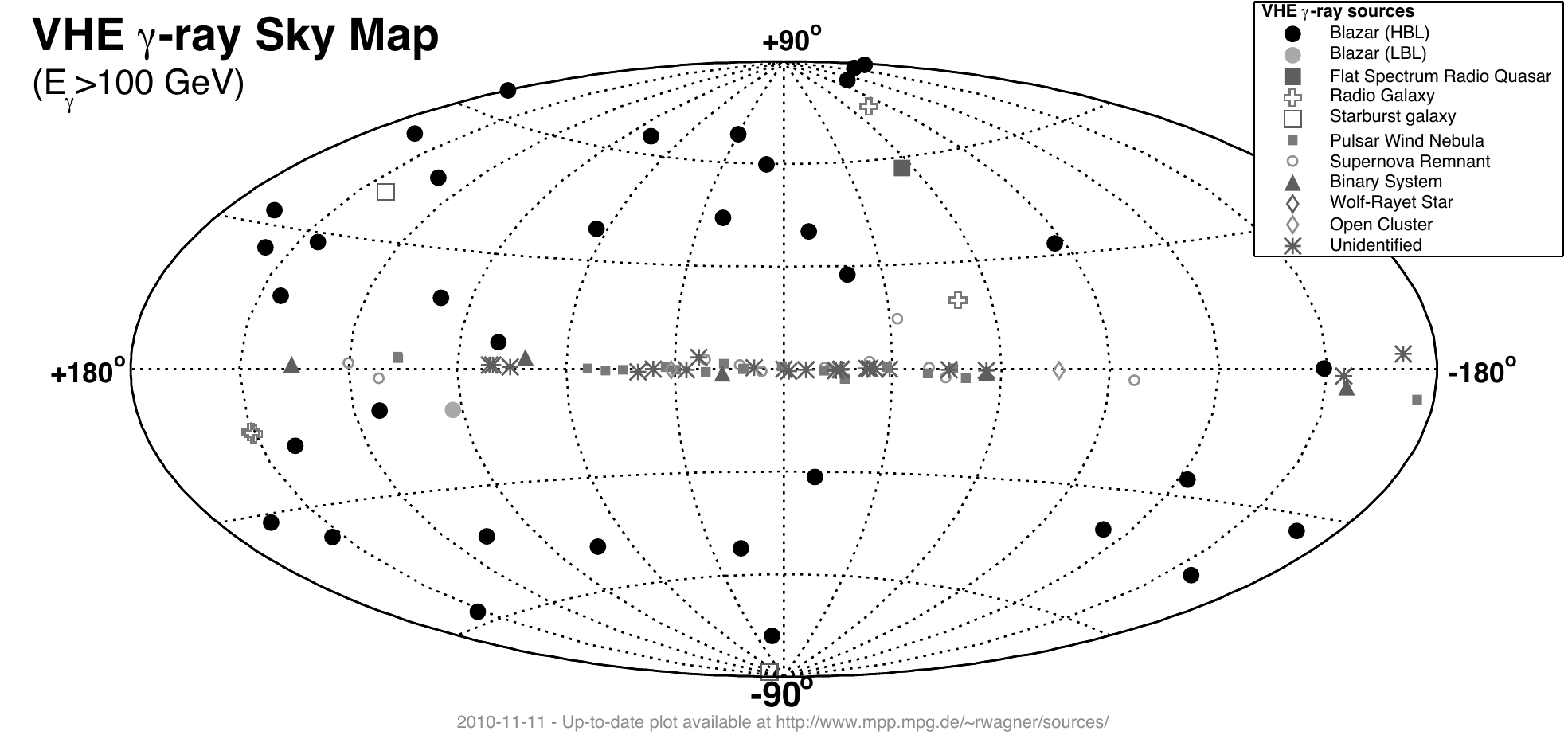}
\caption{\label{fig:wag} Presently known VHE gamma sources (Wagner 2010).}
\end{figure}

Most VHE sources have been
detected and identified in the recent years. For a recent review review on VHE gamma astrophysics, see for example 
(Hinton \& Hofmann 2009, De Angelis et al. 2008). 
When this talk has been presented (September 2010),
more than 100 VHE sources had been detected (see Fig. \ref{fig:wag}); for a comparison, twenty years ago only one source, Crab, had been firmly detected. Among these sources, roughly two thirds are Galactic. 

The recent dramatic growth (by a factor of 10) in the number of known galactic VHE sources (Wagner 2010 and references therein) is largely a consequence of the survey
of the galactic plane conducted with the southern-located detector H.E.S.S. between 2004 and
2007. 
Further galactic sources, accessible
from the northern hemisphere, were subsequently observed with the MAGIC and VERITAS telescopes, and additional sources by H.E.S.S..
Proposed counterparts of such galactic VHE
sources include supernova remnants, PWNe, and accreting binaries.

\begin{table*}
\begin{center}
\begin{tabular}{l c c c c c c c} \hline
Instrument  & \# Tels. & Tel. Area & Field of View & Total Area     & Threshold  & Ang. res. & Sensitivity in 50h\\
                     &           & (m$^2$)  &     ($^{\circ}$)  & (m$^2$) & (TeV)     &  ($^{\circ}$)    &  (\% Crab)\\
\hline
H.E.S.S.    & 4 & 107 & 5 & 428  & 0.1 & 0.06 & 0.7 \\
MAGIC      & 2 & 236 & 3.5 & 472  & 0.05 (0.025) & 0.07 & 0.8\\
       VERITAS  & 4 & 106 & 4 & 424  & 0.1 & 0.07 & 0.7 \\
\hline
\end{tabular}
\caption{Performance of the main Cherenkov telescope arrays.}
\end{center}
           \end{table*}

The extragalactic VHE sources are mostly Active Galactic Nuclei (AGN) of galaxies, blazars in particular; the new discoveries of the recent years are shared democratically among the telescope arrays MAGIC, H.E.S.S. and VERITAS, with MAGIC leading in the detection of far away objects due to a lower
energy threshold.

VHE photons can answer specific questions related to cosmic ray physics and fundamental physics; in particular one can check if photon emission processes continue up to the highest energies. Gamma-rays propagate with very good approximation along straight lines and can be used to locate and study the sources of high energy cosmic rays.
Finally, the highest energies allow testing fundamental physics in an unexplored domain.

There are two main classes of ground based HE gamma detectors: the Extensive Air Shower arrays 
(EAS) and the Cherenkov telescopes.



The EAS detectors, such as MILAGRO and ARGO, are made by a large array of detectors sensitive to charged secondary particles generated by the atmospheric showers.
They have high duty cycle and a large Field of View (FoV), but a low sensitivity.
The energy threshold of EAS detectors is at best in the 0.5 TeV-1 TeV range.
At such energies fluxes are small; thus such detectors need to have large surfaces, of order of~10$^4$~m$^2$.
EAS detectors are possibly provided with a muon detector devoted to hadron rejection; otherwise the discrimination from the background can be done based on the reconstructed shower shape.
The direction of the detected primary particles is computed by taking into account the arrival times of the secondaries, and the angular precision is about 1~degree.
Energy resolution is poor.
\begin{itemize}
\item
The ARGO-YBJ detector at the Tibet site is made of an array of resistive plate counters.
The first results show that ARGO can detect the Crab Nebula with a significance of about 
5\,$\sigma$ in 50~days of observation.
\item
MILAGRO  is a water-Cherenkov based instrument near Los Alamos 
(about 2600~m altitude).
It is made of photomultipliers in water.
It detects the Cherenkov light
produced by the secondary particles of the shower when they pass through the water.
MILAGRO can detect Crab with a significance of about 5\,$\sigma$ in 100~days of 
observation, at a median energy of about 20~TeV.
\end{itemize}


Imaging Atmospheric Cherenkov Telescopes (IACTs), such as H.E.S.S., MAGIC and 
VERITAS, detect the Cherenkov photons produced in air by charged, locally superluminal 
particles in atmospheric showers. The observational technique used by the IACTs is to project the Cherenkov light collected by 
a large optical reflecting surface onto a camera made by an array of photomultiplier tubes 
in the focal plane of the reflector. 
Since they need to operate in dark nights only, due to the faintness of
atmospheric Cherenkov radiation, they have a low duty cycle. The FoV is typically a few degrees only, but they have a high 
sensitivity (they all can detect Crab with a significance of 5$\sigma$ in less than 5 minutes) and a low energy threshold.

Hadronic showers have a different topology with respect to electromagnetic showers, being larger and more subject to fluctuations.
One can thus separate 
showers induced by gamma-rays from the hadronic ones on the basis of the shower shape.

There are three large operating IACTs: H.E.S.S., MAGIC and VERITAS,
 one located in the southern hemisphere and two in the northern 
hemisphere.

\begin{itemize}
\item
The H.E.S.S.\ observatory is composed by four telescopes with surface of 
108~m$^2$ each, working since early 2003, while the first of these telescopes is operating 
since summer 2002.
It is located in the Khomas highlands of Namibia, and it involves several countries, 
Germany and France in particular.
In the future another telescope with a surface of about 600~m$^2$ will be placed in the center 
of the present array.
\item
The MAGIC telescope in the Canary Island of La Palma, is composed of two telescopes with  a diameter of 
17 m and a reflecting surface of 236~m$^2$ each; due to the largest area it reaches the lowest energy threshold.
The collaboration operating MAGIC involves several countries, Germany, Italy, Spain, Finland 
and Switzerland in particular.
Besides the purpose of lowering as much as possible the energy threshold by increasing the 
dish size, the instrument was designed to be able to rapidly slew responding to alerts due to 
transient phenomena (GRBs in particular).
The lightweight construction allows a slewing time of 22~s.
\item
VERITAS  involves Canada, Ireland, the United Kingdom and 
the U.S.A..
The observatory is constituted by an array of four telescopes with a diameter of 12~m and is 
located near Tucson, Arizona.
It is operative since April 2007, but the VERITAS prototype telescope was active since February 2004.
The overall design is rather similar to H.E.S.S..
\end{itemize}
The main characteristics of the operating Cherenkov detectors are summarized in Table 1.%

\section{Insights on the sources of Cosmic Rays}

Among the categories of possible cosmic ray accelerators, several have been studied trying to infer the relation between gammas and charged particles. In the Milky Way in particular, SNRs are since longtime (Baade \& Zwicky 1934)
thought to be possible accelerators to energies of the order of 1 PeV and beyond. The particle acceleration in SNRs is accompanied by production
of gammas due to interactions of accelerated
protons and nuclei with the ambient medium. 

In SNRs with molecular clouds, in particular,
a possible mechanism involves  a source of cosmic rays illuminating clouds at different distances, and generating hadronic showers by $pp$ collisions. This allows to study the generation of cosmic rays by the study of photons coming from $\pi^0$ decays in the hadronic showers. 

\begin{figure}
\begin{center}
\vspace {-2mm}
\includegraphics[angle=0,width=0.65\columnwidth]{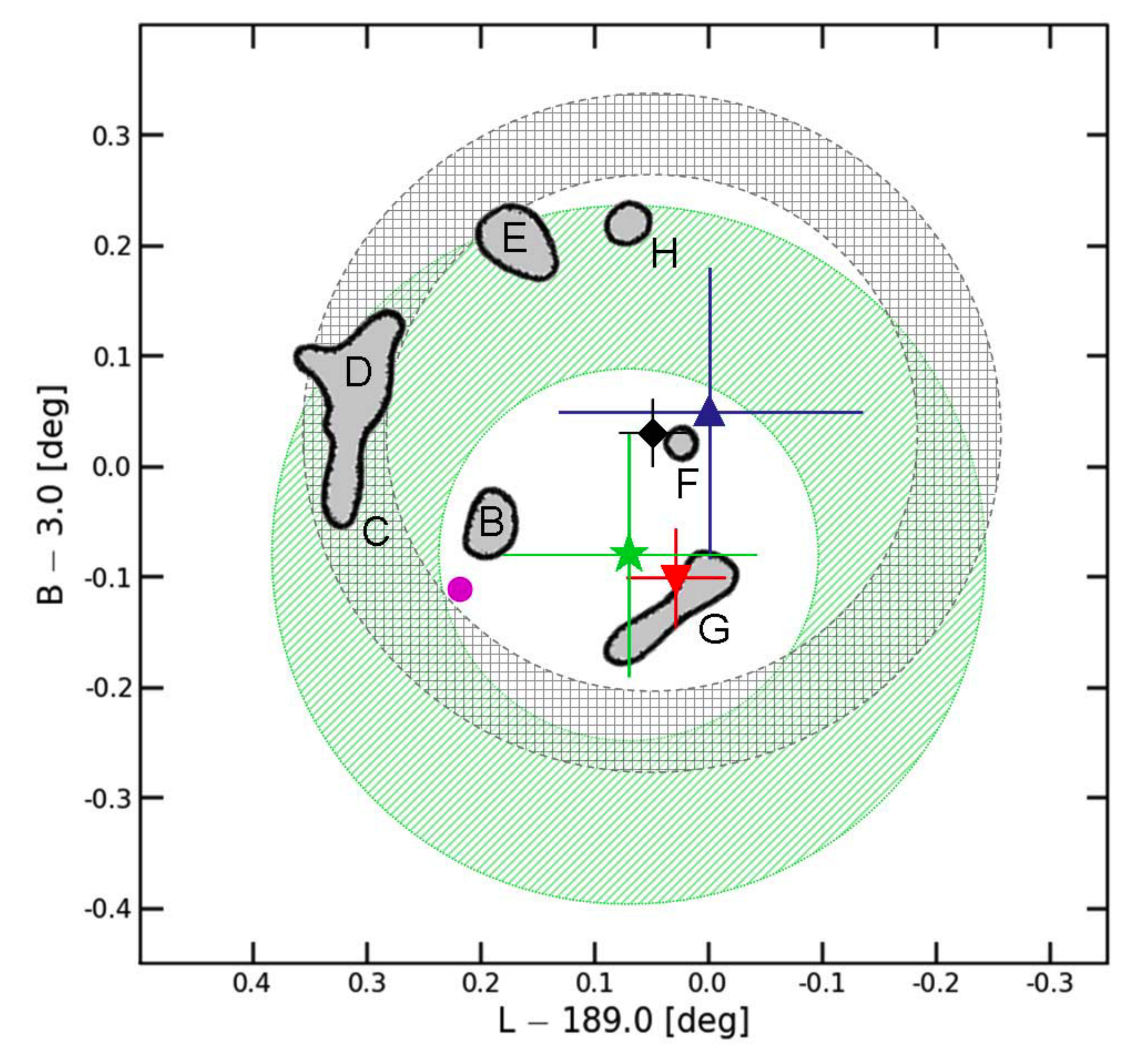}
\end{center}
\vspace {-7mm}
\caption{\label{ic}IC443: centroid of the emission from different gamma detectors. The position from Fermi is marked as a diamond (black), MAGIC as a downwards oriented triangle (red) and is consistent with the molecular cloud G.}
\end{figure}

An example of such a mechanism at work could be IC443 (MAGIC 2007).
Recent results from AGILE (AGILE 2010) and Fermi (Fermi 2010) support the hypothesis. In particular (Fig. \ref{ic}), the centroid of emission by Fermi is significantly displaced from the centroid of emission from MAGIC (which, in turn, is consistent with a later measurement from VERITAS, and with a molecular cloud). The spectral energy distributions also supports a two-component emissions, with a rate of production of primary electrons consistent with the rate of production of protons.

Besides indications from the studies of the morphology, the detection of photons of energies of the order of 100 TeV and above could be a direct indication of production via $\pi^0$ decay, since the emission via leptonic mechanisms should be strongly suppressed at those energies (Aharonian \& Akerlof 2007, and references therein) where the inverse Compton
scattering cross-section enters the Klein-Nishina regime.

As the energetics of SNRs might explain the production of galactic CR, the energetics of AGN might explain the production of CR up to the highest energies.

Recently, The Pierre Auger Collaboration (Auger 2007; Auger 2010) gave
 a marginal indication that the direction of extremely high energy cosmic rays  might be correlated with the spatial distribution of AGNs.

Although the spatial resolution of Cherenkov telescopes is not so good to study the morphology of extragalactic emitters, a recent study of a flare from M87 by the main 
Cherenkov telescopes plus the VLBA radio array (H.E.S.S., MAGIC, VERITAS and VLBA 2009) has shown, based on the VLBA imaging power, that
this AGN accelerates particles to very high energies in the immediate vicinity of its central black hole (less than 60 Schwarzschild radii).

The study of the morphology of galactic sources continues, and is telling us more and more, also in the context of multiwavelength analyses; in the future, the
planned Cherenkov Telescope Array (CTA) will give the possibility to explore the highest energies, and to contribute, together with high energy CR arrays and possibly with neutrino detectors, to the final solution of the CR problem.

\section{The background radiation in the Universe}

Electron-positron $(e^-e^+)$ pair production in the interaction of beam photons off extragalactic background photons is a source of opacity of the Universe to 
$\gamma$-rays.

The dominant process for the absorption is the pair-creation process
$\gamma + \gamma_{\,\texttt{\scriptsize background}} \rightarrow{} e^+ + e^-$, 
for which the cross-section is described by the Bethe-Heitler formula~(Heitler 1960):
\begin{equation}\label{eq.sez.urto}
\sigma(E,\epsilon) \simeq \frac{3\sigma_T}{16} (1-\beta^2) \cdot \left[2 \beta ( \beta^2 -2 )
+ ( 3 - \beta^4 ) \, {\rm ln} \left( \frac{1+\beta}{1-\beta} \right)
\right] 
\end{equation}
where $\sigma_T$ is the Thomson cross section,
$\beta = \sqrt{1-\frac{(m_e c^2)^2}{E\,\epsilon}}$, $m_e$ being the value of the electron mass,
$E$ is the energy of the (hard) incident photon and
$\epsilon$ is the energy of the (soft) background photon.
Notice that only QED, relativity and cosmology arguments are involved in the previous formula.

The cross section  is maximized just above the threshold, $\epsilon \simeq {(500 \,{\rm{GeV}}}/{E})$~eV.
Hence for VHE photons the $\gamma \gamma \rightarrow e^+ + e^-$ interaction becomes 
important with optical/infrared photons -- called extragalactic background light~(EBL), whereas the interaction with the cosmic microwave 
background becomes dominant at $E~\sim$~1~PeV.

The EBL is  the sum of starlight emitted by galaxies throughout their whole cosmic 
history, plus possible additional contributions, like, e.g., light from hypothetical first 
stars that formed before galaxies were assembled.
Therefore, in principle the EBL contains important information both the evolution of baryonic 
components of galaxies and the structure of the Universe in the pre-galactic era.

The probability for a photon of observed energy $E$ to survive absorption along 
its path from its source at redshift $z$ to the observer plays the role of an attenuation 
factor for the radiation flux, and it is usually expressed as $e^{-\tau(E,z)}.$
The coefficient $\tau(E,z)$ is called {\em optical depth}. 

To compute the optical depth of a photon as a function of its observed energy~$E$ and the 
redshift $z$ of its emission one has to take into account the fact that the energy $E$ of a 
photon scales with the redshift~$z$ as $(1+z)$; thus when using 
Eq.~\ref{eq.sez.urto} we must treat the energies as function of~$z$ and evolve 
$\sigma\big(E(z),\epsilon(z),\theta\big)$ for 
$E(z)= (1+z)E$ and $\epsilon(z)=(1+z)\epsilon$,
where $E$ and $\epsilon$ are the energies at redshift $z=0$.
The optical depth is then computed by convolving the photon number density 
of the background photon field with the cross section between the incident $\gamma$-ray and 
the background target photons, and integrating the result over the distance, the scattering 
angle and the energy of the (redshifted) background photon:
\begin{eqnarray}  \label{eq.comp.tau}
\tau(E,z) & = &
\int_{0}^{z} dl(z)\
\int_{-1}^{1}\ d\cos \theta \frac{1-\cos \theta}{2} \; \times \\ & \times &
\int_{\frac{2(m_e c^2)^2}{E(1-\cos\theta)}}^{\infty} d\epsilon(z)\
n_{\epsilon}\big(\epsilon(z),z\big) \ \sigma(E(z),\epsilon(z),\theta) \nonumber
\end{eqnarray}
where
$\theta$ is the scattering angle,
$n_{\epsilon}\big(\epsilon(z),z\big)$ is the density for photons of energy $\epsilon(z)$ at 
the redshift $z$, and $l(z) = c\ dt(z)$ is the distance as a function of the redshift, defined 
by
\begin{equation}
\label{eq:padmanabhan-diff}
\frac{dl}{dz} \ = \ \frac{c}{H_0} 
\frac{1}{(1+z) \sqrt{(1+z)^2 (\Omega_M\,z+1) - \Omega_{\Lambda}\,z(z+2)}} \, .
\end{equation}
In the last formula $H_0$ is the Hubble constant, $\Omega_M$ is the matter density (in units 
of the critical density, $\rho_{\rm c}$) and $\Omega_{\Lambda}$ is the ``dark energy'' density (in 
units of $\rho_{\rm c}$).


The horizon of the observable Universe is expected to rapidly shrink in the very-high-energy 
(VHE) band above $100 \, {\rm GeV}$ as the energy further increases.  Neglecting evolutionary effects for 
simplicity, photon propagation is controlled by the photon mean free path ${\lambda}_{\gamma}(E)$ for 
$\gamma \gamma \to e^+ e^-$, and so the observed photon spectrum $\Phi_{\rm obs}(E,D)$ is related 
to the emitted one $\Phi_{\rm em}(E)$ by 
\begin{equation}
\label{a1}
\Phi_{\rm obs}(E,D) = e^{- D/{\lambda}_{\gamma}(E)} \ \Phi_{\rm em}(E)~.
\end{equation}

Within the energy range in question, ${\lambda}_{\gamma}(E)$ decreases like a power law from the 
Hubble radius $4.2 \, {\rm Gpc}$ around $100 \, {\rm GeV}$ to $1 \, {\rm Mpc}$ around $100 \, {\rm 
TeV}$ (Coppi \& Aharonian 1997). 

\begin{figure}
\begin{center}
\vspace {-6mm}
\includegraphics[width=0.92\columnwidth]{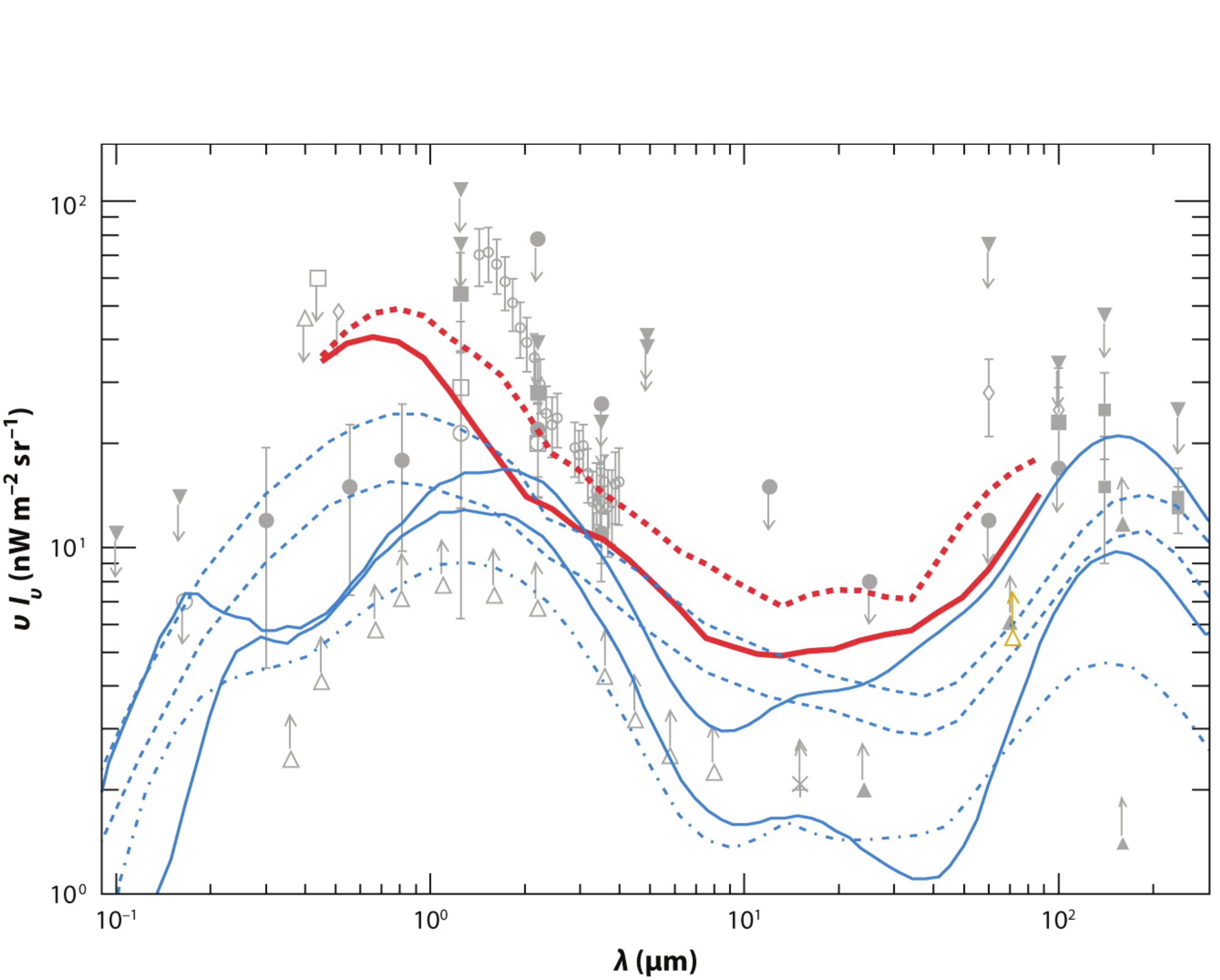}
\caption{\label{fig:density}From (Hofmann \& Hinton 2009): EBL limits obtained from VHE gamma spectra using plausible assumptions for intrinsic spectra (thick red line),
compared to lower limits from direct observations and including recent EBL models. The gray symbols are a collection of EBL
measurements assembled by (Mazin \& Raue 2007); see their Fig. 1 for definitions and references).}
\end{center}
\end{figure}

The attenuation suffered by observed VHE spectra can thus be used to derive constraints on the 
EBL density (Fazio \& Stecker 1971).
First limits on the EBL were obtained in (Stecker et al. 1992), while recent determinations 
from the detection of distant VHE sources are reported in~(Aharonian et al. 2006, Mazin \& Raue 2007).
Fig.~\ref{fig:density} shows the estimated photon number density of the background photons as composed by the radio background, the cosmic microwave background, and 
the infrared/optical/ultraviolet background~(EBL), lower limits from galaxy counts, and upper limits from the observation of distant AGN at VHE.

\subsection{Measurement of cosmological parameters}

Since the 
optical depth depends also on the cosmological parameters (Eq. \ref{eq:padmanabhan-diff}), its determination constrains the 
values of the cosmological parameters if the cosmological emission of galaxies 
and EBL are known, and if only standard processes are at work.

It is shown in~(Blanch \& Martinez 2005) that a determination of $\Omega_M$ and $\Omega_\Lambda$ independent  of the luminosity-distance
relation currently used by the Supernovae 1A observations can be obtained with the observation of distant AGN.

\section{New physics and exotic phenomena}

\subsection{Possible anomalies in photon propagation: Axion-Like Particles}

The energy dependence of the optical depth $\tau$ leads to appreciable modifications of the observed source 
spectrum (with respect to the spectrum at emission) even for small differences in~$\tau$, due to the 
exponential dependence of the attenuation on the optical depth. 
Since the optical depth (and consequently the absoption coefficient) increases with energy, 
the observed flux results steeper than the emitted one.

Thus, eq.~(\ref{a1}) entails that the observed flux is {exponentially} 
suppressed both at high energy and at large distances, so that sufficiently far-away sources become 
hardly visible in the VHE range and their observed spectrum should anyway be steeper than 
the emitted one.

Oservations have {not} detected the behaviour predicted by eq. (\ref{a1}), as shown in (De Angelis et al. 2008) and by the MAGIC observation at 0.5 TeV of the distant blazar 3C279 at $z=0.54$ (MAGIC 2007).

A suggested way out of this difficulty relies upon the modification of the standard Synchro-Self-Compton 
(SSC) emission mechanism. One option invokes strong relativistic shocks (Stecker et al. 2007, Stecker et al. 2008). Another rests 
upon photon absorption inside the blazar (Aharonian et al. 2008). While successful at substantially hardening 
the emission spectrum, all these attempts fail to explain why {only} for the most distant blazars does 
such a drastic departure from the SSC emission spectrum seem to show up.

Another conventional (although less conventional) explanation is the possibility that gamma-ray fluxes from distant AGNs are
enhanced by relatively nearby production by interactions of primary cosmic rays emitted from the same source
(Essey et al. 2010). 

An alternative proposal -- usually referred to as the DARMA scenario -- can be summarized as follows. 
Photons can oscillate into a very 
light spin-zero Axion-Like Particle (ALP)  due to cosmic magnetic 
fields ((De Angelis, Roncadelli \& Mansutti 2007) or in the source (Hooper \& Serpico 2007). Once ALPs are produced 
close enough to the source, they travel {unimpeded} throughout the Universe and can convert back to photons 
before reaching the Earth. Since ALPs do not undergo EBL absorption, the  {effective} photon mean free path 
${\lambda}_{\gamma , {\rm eff}} (E)$ gets {increased} so that the observed photons cross a distance in excess 
of ${\lambda}_{\gamma}(E)$. Correspondingly, eq. (\ref{a1}) becomes
\begin{equation}
\label{a1bis}
\Phi_{\rm obs}(E,D) = e^{- D/{\lambda}_{\gamma, \,{\rm eff}}(E)} \ \Phi_{\rm em}(E)~,
\end{equation}
from which we see that even a {slight} increase of ${\lambda}_{\gamma,\, {\rm eff}}(E)$ with respect to ${\lambda}_{\gamma}(E)$ can give rise to a 
{huge} enhancement of the observed flux.

\subsection{Probing Special Relativity}

 Besides the general interest in verifying Special Relativity (SR), several theories 
(Mattingly 2005 and references therein) predict Lorentz-invariance violations (LIV), for example via modifications of 
the propagation of energetic particles. Dispersive effects due to a non-trivial refractive 
index can be induced, for example, by Quantum Gravity (QG) effects.

\subsubsection{Does $c$ depend on the photon energy?}

The constancy of the speed of light, $c$, is one of the postulates of SR; 
it has been extensively tested in the recent years against a possible dependence on the photon 
energy.

The dependence of the speed of light on the energy $E$ of the photon is usually parametrized as
\begin{equation}
c' = c \left[ 1 \pm \left( \frac{E}{E_{s1}} \right) 
\pm \left( \frac{E}{E_{s2}} \right)^2  \pm ...
\right] \, .
\end{equation}
The energy scales $E_{s1}$, $E_{s2}$ which parametrize the LIV in the above expression 
are usually expressed in units of the Planck mass, $M_P \simeq 1.22 \times 10^{19}$~GeV 
(natural units with $\hbar$ = c =1 are employed throughout). If the linear term dominates, 
the above expression reduces to
\begin{equation}\label{amelino}
c' = c \left[ 1 \pm \left( \frac{E}{E_{s1}} \right) 
\right] \, .
\end{equation}

A favored way to search for such a dispersion relation is to compare the arrival times 
of photons of different energies arriving on Earth from pulses of distant astrophysical 
sources.
The greatest sensitivities may be expected from sources 
with short pulses, at large distances or redshifts $z$, of photons observed over a large 
range of energies. In the past, studies have been made of emissions from pulsars, 
$\gamma$-ray bursts (GRBs) and active galactic nuclei (AGN).

Recently, the new Cherenkov imaging detectors and the Fermi/LAT gamma-ray telescope 
gave the opportunity to test this hypothesis at an unprecedented precision.

 Data from Cherenkov telescopes and from Fermi do not allow to draw a clear picture for the moment.
 A correlation seen by MAGIC experiment (MAGIC \& Ellis et al. 2008) in a flare by Markarian 501
  between the photon 
energy and the arrival time has not been confirmed in further flares observed by H.E.S.S. and in GRBs by Fermi.
 
The figures of merit for the sensitivity to effects related to new physics are the energy 
span $\Delta E$ and the redshift $z$. Especially for $\Delta E$ new generation Cherenkov 
telescopes will substantially improve the panorama.

\subsubsection{Energy thresholds effects in absorption processes}

A powerful tool to investigate Planck-scale
departures from Lorentz symmetry could be provided by certain types of energy thresholds in the
pair production process $\gamma_{VHE} \gamma_{EBL} \to e^+ e^-$ of gamma-rays from cosmological sources.
This would affect the optical depth.

In a collision between a soft photon of energy $\epsilon$ and a high-energy photon of energy $E$, an 
electron-positron pair could be produced only if $E$ is greater than the threshold energy $E_{th}$,
which depends on $\epsilon$ and $m_{e}^2$.
Using a dispersion relation (Amelino-Camelia et al. 1998) of the form
\begin{equation}
m^2 ~\simeq~ E^2 - {\bf p}^2 + \xi {\bf p}^2 ({E^n}/{E_{\rm p}^n})
\label{eq_disp}
\end{equation}
with real $\xi$ and $n$ integer ($>0$), one obtains, for $n = 1$ and unmodified law of energy-momentum 
conservation, that for a given soft-photon energy $\epsilon$, the process $\gamma \gamma \to e^+ e^-$ 
is allowed only if $E$ is greater than a certain threshold energy $E_{th}$ which depends on $\epsilon$ 
and $m_{e}^{2}$:
\begin{equation}
E_{th}\epsilon + \xi ({E^3_{th}}/{8E_{p}}) \simeq m_{e}^{2}.
\end{equation}
The $\xi~\rightarrow~0$ limit corresponds to the special-relativistic result $E_{th}$ = $m_{e}^{2}$ / $\epsilon$.
For $|\xi|~\sim~1$ and sufficiently small values of $\epsilon$ (and correspondingly large values of $E_{th}$ ) 
the Planck-scale correction cannot be ignored.

This provides an opportunity for pure-kinematics tests. As an example, a 10~TeV photon and a 0.03~eV photon 
can produce an electron-positron pair according to ordinary special-relativistic kinematics, but they cannot 
produce a $e^+ e^-$ pair according to the dispersion relation in Eq.~\ref{eq_disp}, with $n=1$ and $\xi \sim -1$.
The non-observation of EeV gamma-rays has already excluded a good
part of the parameter range of terms suppressed to first and
second order in the Planck-scale (Galaverni \& Sigl 2008).

The situation for positive $\xi$ is somewhat different, because a positive $\xi$ decreases the energy requirement 
for electron-positron pair production.

and certainly within a few years dramatic improvements will occur. This means that this strategy of 
measurements and analysis will possibly take us at the Planck-scale sensitivity and beyond.

\subsection{Dark Matter}

Evidence for departure of cosmological motions from the predictions of Newtonian dynamics 
based on visible matter, are well established 
-- from galaxy scales to galaxy-cluster scales to cosmological scales. 
Such departures could be interpreted as due to the presence of an undectected matter (Zwicky 1930), called Dark Matter (DM).  

A remarkable
agreement of a diverse set of astrophysical data indicates that the energy budget of DM could be ~26\% of the 
total energy of the Universe, compared to some 4\% due  to ordinary matter.

DM particle candidates should be weakly interacting with ordinary matter (and hence neutral). 
The theoretically favored ones, which are heavier than the proton, are dubbed weakly interacting 
massive particles (WIMPs). WIMPs should be long-lived enough to have survived from their 
decoupling from radiation in the early universe into the present epoch. Except for the neutrino, 
which is the only DM particle known to exist within the Standard Model (SM) of elementary particles 
(with a relic background number density of $\sim$50~ cm$^{-3}$ for each active neutrino species) 
but which is too light ($m_\nu < 1$~eV) to contribute significantly to $\Omega_m$ given 
the current cosmological model, WIMP candidates have been proposed within theoretical 
frameworks mainly motivated by extensions of the SM (e.g., the 
R-parity conserving supersymmetry, SUSY). Among current WIMP candidates, 
the neutralino $\chi,$ which is the lightest SUSY particle, is the most popular candidate. Its relic 
density is compatible with {\it W}MAP bounds, if its mass is of the order of 10-100 GeV.

WIMPs could be detected directly, via elastic scattering vith targets on Earth, or indirectly, by 
their self-annihilation products (neutralinos are Majorana particles) in high-density DM environments, or by decay products. 

\subsubsection{VHE gamma signatures}

DM self-annihilation can generate 
$\gamma$-rays through several processes. Most distinctive are those that result in mono-energetic 
spectral lines, $\chi\chi$$\rightarrow$$\gamma\gamma$, $\chi\chi$$\rightarrow$$\gamma$$Z$ or 
$\chi\chi$$\rightarrow$$\gamma$$h$. However, in most models the processes only take place through 
loop diagrams; hence the cross sections for such final states are quite suppressed, and the lines 
are weak and experimentally challenging to observe. A continuum $\gamma$-ray spectrum can also be 
produced through the fragmentation and cascades of most other annihilation products. The resulting 
spectral shape depends on the dominant annihilation modes, whereas 
the normalization depends on the WIMP's velocity-averaged annihilation cross section as well 
as on the DM density profile. 

Once an astroparticle model has been established, the main uncertainties 
are of astrophysical nature. Superposed to any VHE emission from the decaying DM (cosmological, 
non-baryonic signal), galaxies can display a VHE emission from astrophysical sources associated with 
the visible matter distribution (astrophysical, baryonic signal). The ratio of the former to the 
latter is maximized in small, low-luminosity galaxies, because the dark-to-visible mass ratio 
as well as the central DM density increase with decreasing luminosity. Clearly, 
distance dilution of the signal opposes detection, so galaxies candidate for indirect DM detection 
should be chosen among nearby objects. In conclusion, the best obserational targets for DM detection 
are the Milky Way dwarf spheroidal galaxies (e.g., Draco, Sculptor, Fornax, Carina, Sextans, Ursa 
Minor). A further issue, stemming from the $\rho^2$ dependence (as a result of annihilation) of the 
normalization integral of the $\gamma$-ray emission, concerns the shape of the inner halo profile, 
i.e. whether the latter is cuspy or cored. Cuspy profiles are produced in cosmological N-body 
simulations of halo formation, whereas cored profiles are suggested by the 
measured rotation curves of disk galaxies. 

These considerations (and uncertainties) have been incorporated in detailed predictions of the 
$\gamma$-ray flux expected from the annihilation of the neutralino pairs. Outlooks for VHE neutralino 
detection in Draco by current IACTs are not very promising: for a neutralino mass $m_\chi$$=$100~GeV 
and a variety of annihilation modes, and in the favorable case of a maximal (cuspy) inner halo profile, 
VHE detection (by MAGIC in 40hr observation) can occur if average value of the neutralino's cross 
section times velocity is $<$$\sigma$$v$$> > $10$^{-25}$~cm$^3$s$^{-1}$, which is somewhat larger 
than the maximum value for a thermal relic with a density equal to the measured (cold) DM density (but 
may be fine for non-thermally generated relics) in the allowed SUSY parameter space. The prospects are 
better in the Fermi energy range. 

No evidence of DM annihilation $\gamma$-rays has been unambiguously claimed so far. 

Clearly, IACT(+ Fermi/{\it LAT}, possibly) positive detections from the direction to known dark halos, all 
characterized by the same spectral signatures, would be seen as a powerful indication that neutralino 
decay has been detected -- and hence the nature of DM has been unveiled. 

The possibility to use diffuse VHE photons (in a similar way as GeV photon detectors do) seems, instead, remote for VHE.


\subsubsection{Electron (positron) signatures}

Gamma detectors (Cherenkov in particular) can also study the electron yield.

Claims by the ATIC balloon experiment (ATIC 2008) of a peak in the yield of electrons (plus positrons) have been recently not confirmed by Fermi (Fermi 2010)
and H.E.S.S. (HESS 2008).
The study of the spectrum of electrons and positrons at high energy is anyway of extreme interest, and Cherenkov telescopes can contribute in the VHE range.

Recently, PAMELA has observed
an anomalous positron abundance in cosmic
radiation (PAMELA 2009). The positron over electron fraction
appears to rise at energies from
10 GeV to 100 GeV;
no obvious antiproton excess is seen for the
same energy range. Many suggestions have been
made to explain the positron excess at PAMELA.
Among different approaches, DM annihilation
is especially interesting.

The study of the ratio $e^+/e^-$ at higher energies would give useful information;
also an upper limit could constrain models.
Separation of electrons from positrons in Cherenkov telescopes can be achieved by IACTs (Colin et al. 2009).
The Moon produces a 0.5$^\circ$-diameter hole in the isotropic CR flux, which is shifted by the Earth magnetosphere depending on the momentum and charge of the particles. Below a few TeV, the positron and electron shadows are shifted by more than $0.5^\circ$ each side of the Moon due to the geomagnetic field.

\subsection{New heavy particles: top-down mechanisms}

Extragalactic gamma-ray emission could originate in decays of exotic particles in the early
Universe. The energy spectrum of this component should be different from the AGN contributions
(Kamionkowski 1995, Willis 1996). Bounds on long-lived relics have been derived using EGRET and
COMPTEL observations of the diffuse $\gamma$-ray background (Kribs \& Rothstein 1997). Many models predict 
long-lived relics that may or may not be dark matter candidates. Long lifetimes for heavy relics,
larger than the age of the Universe, may arise in models with symmetry breaking at short distances.
Examples of such models are technibaryons in technicolor models or R-parity violating SUSY.

\section{The future}
The future is bright for VHE gamma astrophysics. 

In the next few years, thanks to foreseen upgrades of H.E.S.S. and MAGIC, the morphology of Galactic sources will be better studied. In the medium term,
the HAWC project (HAWC 2010) will allow a serendipitous study of sources, extended sources in particular.

The number of VHE sources, now of the order of 100, should increase by a factor of 2-3 thanks to the foreseen short- and medium- term projects.
The advent of CTA (CTA 2010) should further increase the statistics and the diversity of known Galactic and extragalactic sources.  The detection of DM is a major objective of the field, although the phase space is narrow. A quantitative understanding of Galactic CR origin seems close.

And, of course, a better knowledge of the Universe in an unexplored domain opens a wide spectrum of possible surprises related to completely new phenomena.

\begin{acknowledgements}
Thanks to the organizers, and in particular to J. Poutanen, F. Aharonian and
I. Usoskin, for the invitation. I thank B.  De Lotto, M.  Persic and M.  Roncadelli for providing material for this note.
\end{acknowledgements}

\end{document}